\documentclass[a4paper,12pt]{article}  
\usepackage{mathptmx}
\usepackage{amssymb}
\usepackage{amsmath}
\usepackage{graphicx}
\usepackage[dvipdfm]{hyperref}

\newdimen\einr
\def\abs#1{\par\hangafter=1\hangindent=\einr
  \noindent\hbox to\einr{\ignorespaces#1\hfill}\ignorespaces} 

\newtheorem{theorem}{Theorem}

\def\reals{{\mathbb R}}

\def\proof{\noindent{\em Proof.\enspace}}

\def\endproof{\hfill\strut\nobreak\hfill\tombstone\par\medbreak}
\def\tombstone{\hbox{\lower.4pt\vbox{\hrule\hbox{\vrule
  \kern7.6pt\vrule height7.6pt}\hrule}\kern.5pt}}

\def\9{{\mathbb 0}}
\def\1{{\bf 1}}
\def\T{^{\top}}

\parindent\einr

\title{%
Rank-1 Games With Exponentially Many Nash Equilibria
}

\author{%
Bernhard von Stengel%
\thanks{Department of Mathematics,
London School of Economics, London WC2A 2AE, United Kingdom.
Email: stengel@nash.lse.ac.uk}
}

\date{November 11, 2012}

\begin{document}
\maketitle
\begin{abstract}
\noindent
The rank of a bimatrix game $(A,B)$ is the rank of the
matrix $A+B$.
We give a construction of rank-1 games with exponentially
many equilibria, which answers an open problem by Kannan and
Theobald (2010).

\strut


\strut


\strut


\end{abstract}


\noindent
Finding a Nash equilibrium of a bimatrix game is a
PPAD-complete problem (Chen and Deng, 2006).
For that reason, classes of bimatrix games where a Nash
equilibrium can be found more easily are of some interest.
An equilibrium of a zero-sum game $(A,B)$ where $A+B$ is the
all-zero matrix can be found in polynomial time by solving a
linear program.
As a generalization, Kannan and Theobald (2010) defined the
{\em rank} of a bimatrix game $(A,B)$ as the rank of the
matrix $A+B$, and give a polynomial-time algorithm to find
an approximate equilibrium of a game of fixed rank.  
They asked (Open Problem~9) if a rank-1
game may possibly have only a polynomial number of Nash
equilibria.
This is not the case, according to the following theorem.

\begin{theorem}
\label{t-expo}
Let $p>2$ and let $(A,B)$ be the $n\times n$ bimatrix game
with entries of $A$ 
\begin{equation}
\label{aij}
a_{ij}=
\begin{cases}
2p^{i+j} & \hbox{if } j>i\\
p^{2i} & \hbox{if } j=i\\
0 & \hbox{if } j<i\\
\end{cases}
\end{equation}
for $1\le i,j\le n$, and $B=A\T$.
Then $A+B$ is of rank $1$, and $(A,B)$ is a nondegenerate
bimatrix game with $2^n-1$ many Nash equilibria.
\end{theorem}

\proof
By (\ref{aij}), $A+B=\alpha\beta\T$ with the $n$ components of
the column vectors $\alpha$ and $\beta$ defined by
$\alpha_i=p^i$ and $\beta_j=2p^j$ for $1\le i,j\le n$, so
$A+B$ is of rank~1.

Let $y$ be any mixed strategy of the column player and let
$S$ be the support of~$y$, that is, $S=\{j\mid y_j>0\}$.
Consider any row~$i$ and let $T=\{j\in S\mid j>i\}$.
Then, because $A$ is upper triangular, the expected payoff
against $y$ in row $i$ is
\begin{equation}
\label{Ayi}
(Ay)_i=a_{ii}y_i+\sum_{j\in T}a_{ij}y_j\,.
\end{equation}
Suppose $i\not\in S$.
If $T$ is empty, then $(Ay)_i=0<(Ay)_1$, otherwise 
let $t=\min T$ and note that for $j\in T$ we have
$a_{ij}=2p^{i+j}<p^{1+i+j}\le p^{t+j}\le a_{tj}$, so 
$(Ay)_i<(Ay)_t$.
Hence, no row~$i$ outside $S$ is a best response to $y$.
Similary, because the game is symmetric, any column that is
a best response to a strategy $x$ of the row player belongs
to the support of~$x$.
So no mixed strategy has more pure best responses than the
size of its support, that is, the game is nondegenerate (von
Stengel, 2002).
Moreover, if $(x,y)$ is a Nash equilibrium of $(A,B)$, then
$x$ and $y$ have equal supports.

For any nonempty subset $S$ of $\{1,\ldots,n\}$, we
construct a mixed strategy $y$ with support~$S$ so that
$(y,y)$ is a Nash equilibrium of $(A,B)$.
This implies that the game has $2^n-1$ many Nash equilibria,
one for each support set $S$.
The equilibrium condition holds if $(Ay)_i=u$ for $i\in S$
with equilibrium payoff~$u$, because then $(Ay)_i<u$ for
$i\not\in S$ as shown above.
We start with $s=\max S$, where $(Ay)_s=a_{ss}y_s=u$, by
fixing $u$ as some positive constant (e.g., $u=1$), which
determines $y_s$.
Once $y_i$ is known for all $i\in S$ (and $y_i=0$ for
$i\not\in S$), we scale $y$ and $u$ by multiplication with
$1/\sum_{i\in S}y_i$ so that $y$ becomes a mixed strategy.
Assume that $i\in S$ and $T=\{j\in S\mid j>i\}\ne\emptyset$
and assume that $y_k$ has been found for all $k$ in $T$ so
that $(Ay)_k=u$ for all $k$ in $T$, which is true for
$T=\{s\}$.
Then, as shown above,
$\sum_{j\in T}a_{ij}y_j<\sum_{j\in T}a_{tj}y_j=(Ay)_t=u$ for
$t=\min T$, so $y_i$ is determined by $(Ay)_i=u$ in
(\ref{Ayi}), and $y_i>0$.
By induction, this determines $y_i$ for all $i$ in $S$, and
after re-scaling gives the desired equilibrium strategy~$y$.  
\endproof

Adsul, Garg, Mehta, and Sohoni (2011) showed how to find in
polynomial time an exact Nash equilibrium of a rank-1 game,
which is of the form $(A,-A+\alpha\beta\T)$ with suitable
column vectors $\alpha\in\reals^m$ and $\beta\in\reals^n$.
They proved that a mixed strategy pair
$(x,y)$ is a Nash equilibrium of this game if and only if
for some suitable real $\lambda$ the equation
$x\T\alpha=\lambda$ holds and $(x,y)$ is a Nash equilibrium
of the game $(A,-A+\1\lambda\beta\T)$, where $\1$ is the
all-one vector; this equilibrium can be found as the
solution to a linear program parameterized by~$\lambda$.
Their algorithm uses binary search for $\lambda$ combined
with solving the parameterized LP.

The exponential number of Nash equilibria of the game in
Theorem~\ref{t-expo} shows that the path that follows the
solutions of the parameterized LP with parameter~$\lambda$
has an exponential number of intersections with the
hyperplane defined by $x\T\alpha=\lambda$.
Hence, that path has exponentially many line segments.
Murty (1980) describes a parameterized LP with such an
exponentially long path of length $2^n$.
His LP is of the form
\begin{equation}
\label{murty}
\hbox{maximize }{c\T z} 
\quad\hbox{subject to }{A z}\ge b+\1\lambda,~~
z\ge0
\end{equation}
with $A$ as in (\ref{aij}) with $p=1$, and the vectors $c$
and $b$ in $\reals^n$ given by $c_j=4^{n-j}$ and
$b_i=-2^{n-i}$ for $1\le i,j\le n$.
The payoffs for the game in Theorem~\ref{t-expo} have been
inspired by Murty's example, but are not systematically
constructed from it;
at the point of this writing, it is not even clear how to
get a game with that many equilibria from Murty's result.

For specific instances of the game in Theorem~\ref{t-expo}
one can choose $p=3$ or $p=4$ in (\ref{aij}) and divide all
payoffs by $p^2$ (or let the rows and columns be numbered
from $0$ to $n-1$ rather than $1$ to $n$).
In the construction of mixed strategies $y$ with support $S$
described in the proof, starting with $u=p^s$ then gives
integer values for $y_i$ for $i\in S$ which are afterwards
re-scaled.
Verifying the equilibria of these games was aided by the
webpage of Savani (2012).

The number of $2^n-1$ Nash equilibria of an $n\times n$
bimatrix game is large, the same as that of the coordination
game where both players have the identity matrix (which has
maximal rank).
Quint and Shubik (1997) even conjectured this to be the
maximum possible number (always considering nondegenerate
games), which is true for $n\le 4$ (Keiding, 1997; McLennan
and Park, 1999).
However, this conjecture was refuted by von Stengel (1999)
who constructed a $6\times 6$ game with 75 equilibria, and
more generally $n\times n$ games with asymptotically more
than $2.4^n$ equilibria.
Quint and Shubik (2002) showed that a game $(A,A)$ where
both players have identical payoffs has at most $2^n-1$ 
equilibria.
A {\em symmetric} game $(A,A\T)$ of size $n\times n$, as
considered in Theorem~\ref{t-expo}, has at most $2^n-1$
symmetric equilibria, because an equilibrium is uniquely
determined by the pair of supports for the two strategies.
However, the number of possibly nonsymmetric equilibria of a
symmetric game is not bounded by $2^n-1$, as the following
simple argument based on a standard symmetrization shows.
Suppose $(A,B)$ is an $n\times n$ bimatrix game with
positive payoff matrices and more than $2^n$ equilibria,
and let
$C=\left(\begin{matrix}0 & A\\ B\T & 0\end{matrix}\right)$.
Then for any {\em pair} of equilibria $(x,y)$, $(x',y')$ of
$(A,B)$, one obtains an equilibrium
$((\hat x, \hat y'),(\hat x',\hat y))$ of $(C,C\T)$ where
$\hat x$, $\hat x'$, $\hat y$, and $\hat y'$ are scaled
versions of $x$, $x'$, $y$, and $y'$, respectively, so that
the respective optimal payoffs of $A\hat y$ and $B\T\hat x'$
coincide, and similarly those of $B\T\hat x$ and $A\hat y'$.
Then $(C,C\T)$ is of size $2n\times 2n$ and has more than
$(2^n)^2$ many equilibria, as claimed.

Hence, it is an open question if there are nondegenerate
$n\times n$ games of rank~1 with more than $2^n$ many Nash
equilibria.

\section*{References}
\frenchspacing
\parindent=-\einr\advance\leftskip by\einr
\parskip=.6ex plus1pt minus1pt
\small
\hskip\parindent
Adsul, B., J. Garg, R. Mehta, and M. Sohoni (2011),
Rank-1 bimatrix games: a homeomorphism and a polynomial time algorithm.
Proc. 43rd 
STOC, 195--204. 

Chen, X., and X. Deng (2006),
Settling the complexity of two-player Nash equilibrium.
Proc. 47th FOCS, 261--272.

Kannan, R., and T. Theobald (2010),
Games of fixed rank: a hierarchy of bimatrix games.
Economic Theory 42, 157--173.

Keiding, H. (1997),
On the maximal number of Nash equilibria in an $n\times n$
bimatrix game.
Games and Economic Behavior 21, 148--160.  


McLennan, A., and I.-U. Park (1999),
Generic $4\times 4$ two person games have at most 15 Nash
equilibria.
Games and Economic Behavior 26, 111--130.  

Murty, K. G. (1980),
Computational complexity of parametric linear programming.
Mathematical Programming 19, 213--219.

Quint, T., and M. Shubik (1997),
A theorem on the number of Nash equilibria in a bimatrix game.
International J. Game Theory 26, 353--359.

Quint, T., and M. Shubik (2002),
A bound on the number of Nash equilibria in a coordination game. 
Economics Letters 77, 323--327. 

Savani, R. (2012), Solve a bimatrix game.
Interactive webpage at
\url{http://banach.lse.ac.uk}.

von~Stengel, B. (1999),
New maximal numbers of equilibria in bimatrix games.
Discrete and Computational Geometry 21, 557--568.

von~Stengel, B. (2002),
Computing equilibria for two-person games.
Chapter 45, Handbook of Game Theory, Vol.~3,
eds. R. J. Aumann and S. Hart, North-Holland, Amsterdam,
1723--1759.

\end{document}